\documentclass[aps,prl,10pt,twocolumn,longbibliography,amsmath,amssymb,floatfix,superscriptaddress]{revtex4-2}

\usepackage{graphicx}
\graphicspath{{fig_shaped/}}
\usepackage{dcolumn}
\usepackage{bm}
\usepackage{xcolor}
\usepackage{txfonts}
\usepackage{mathtools}
\usepackage{slashed}
\usepackage{braket}
\usepackage{tensor}
\usepackage{siunitx}
\usepackage{booktabs}
\usepackage{multirow}
\usepackage{microtype}
\usepackage[caption=false]{subfig}
\usepackage{hyperref}

\begin{document}

\title{Photon Orbital Angular Momentum Control by Electron Wavepackets in Nonlinear Compton Scattering}
\author{Zheng-Yang Zuo}
\thanks{These authors contributed equally to this work. }
\affiliation{Department of Nuclear Science and Technology, Xi'an Jiaotong University, Xi'an 710049, China}
\author{Peng-Pei Xie}
\thanks{These authors contributed equally to this work. }
\affiliation{Department of Nuclear Science and Technology, Xi'an Jiaotong University, Xi'an 710049, China}
\author{Xiang-Nan Shi}
\affiliation{Department of Nuclear Science and Technology, Xi'an Jiaotong University, Xi'an 710049, China}

\author{Yan-Fei Li}\email{liyanfei@xjtu.edu.cn}    
\affiliation{Department of Nuclear Science and Technology, Xi'an Jiaotong University, Xi'an 710049, China}

\begin{abstract}
Photon orbital angular momentum (OAM) generated in nonlinear Compton scattering has attracted considerable interest as a route toward vortex $\gamma$-ray sources. Existing theories describe photon OAM primarily through angular-momentum transfer involving structured incident particles and laser fields, while the role of electron-wavepacket  has remained unexplored. Here, we develop a general analytical theory of nonlinear Compton scattering for arbitrarily shaped electron wave packets and demonstrate that the Fourier spectrum of the transverse electron wave packet directly determines the OAM spectrum of the emitted photons through a generalized angular-momentum selection rule. Our theory unifies Gaussian wave packets, vortex electrons, and arbitrarily shaped electron states within a single framework, revealing the transverse Fourier structure of electron wave packets as a fundamental degree of freedom governing photon orbital angular momentum and enabling deterministic engineering of photon OAM distributions.
\end{abstract}

\maketitle
Vortex photons carrying orbital angular momentum (OAM) have attracted considerable attention owing to their unique spatial structure and additional degree of freedom beyond energy, linear momentum, and polarization\cite{BLIOKH20151,knyazev2018beams,erhard2018twisted,ivanov2022promises,forbes2025progress,fang2025ultrafast}. Their helical wavefronts enable applications ranging from high-dimensional quantum information and quantum-state encoding\cite{qiu2023remote,scarfe2025high} to light--matter interactions with modified selection zrules\cite{lange2022excitation,PhysRevLett.131.202502,liu2026vortex}. Extending structured light into the high-energy regime would provide a powerful tool for investigating angular-momentum-resolved strong-field and quantum electrodynamic (QED) processes. Consequently, the generation and manipulation of vortex $\gamma$ photons have become an important objective in modern strong-field physics.

In contrast to optical wavelengths, generating high-energy vortex photons remains highly challenging due to the lack of efficient phase-modulation techniques at short wavelengths \cite{Beijersbergen1994,Heckenberg1992,Forbes2016}. With the rapid development of ultraintense laser facilities and relativistic electron sources, Compton scattering has emerged as one of the most promising routes toward vortex $\gamma$-ray generation. Early studies demonstrated that linear Compton backscattering can transfer the OAM of incident vortex photons to scattered photons\cite{PhysRevLett.106.013001,jentschura2011compton}.  Recently, quantum-theoretical studies based on strong-field QED demonstrated that nonlinear Compton scattering (NCS) in intense circularly polarized laser fields can directly generate vortex $\gamma$ photons through multiphoton absorption processes\cite{ababekri2024vortex,guo2024generation,bu2024generation,liao2025all}. Subsequent studies further explored the manipulation of photon angular momentum using radiation-reaction effects\cite{chen2018gamma} and two-color counter-rotating laser fields\cite{PhysRevLett.134.153802}. Notably, the recent experimental observation of sub-MeV vortex $\gamma$ photons generated by all-optical inverse Compton scattering provides evidence for the feasibility of this mechanism\cite{wei2026experimental}. Despite these advances, precise control of the photon OAM spectrum in NCS remains elusive. Existing approaches determine the accessible OAM channels primarily through the angular momentum carried by the incident particles, laser fields, and multiphoton absorption processes, leaving little flexibility once the interaction configuration is specified.

More fundamentally, a relativistic electron participating in a scattering process is a quantum wave packet rather than a momentum eigenstate. Beyond its average momentum or angular momentum, an electron wave packet possesses a transverse amplitude and phase distribution that constitutes an intrinsic quantum degree of freedom. Recent advances in free-electron quantum optics have enabled preparation and manipulation of electron wave packets using holographic phase masks, optical near fields, integrated photonic structures, and laser--electron interactions\cite{McMorran2011,Feist2015,Henke2021,tsarev2023nonlinear,garcia2021optical,chirita2022transverse,ruimy2025free}. Such wave-packet engineering has been shown to modify quantum radiation through coherence and interference effects\cite{wong2021control,lim2023quantum}, suggesting that the internal structure of an electron wave packet may itself participate in determining scattering observables.

This raises a fundamental question. Can the internal structure of an electron wave packet itself determine the orbital angular momentum of emitted photons? Existing theories describe photon OAM in NCS through angular-momentum conservation associated with multiphoton absorption, where Gaussian wave packets and vortex electrons correspond to particular initial states. However, a general electron wave packet contains far richer information encoded in its transverse Fourier spectrum than a single angular-momentum quantum number. Whether this internal structure provides a new physical degree of freedom governing the OAM spectrum of emitted photons has remained unexplored.

\begin{figure}[tbp]
    \centering
    \includegraphics[width=0.48\textwidth]{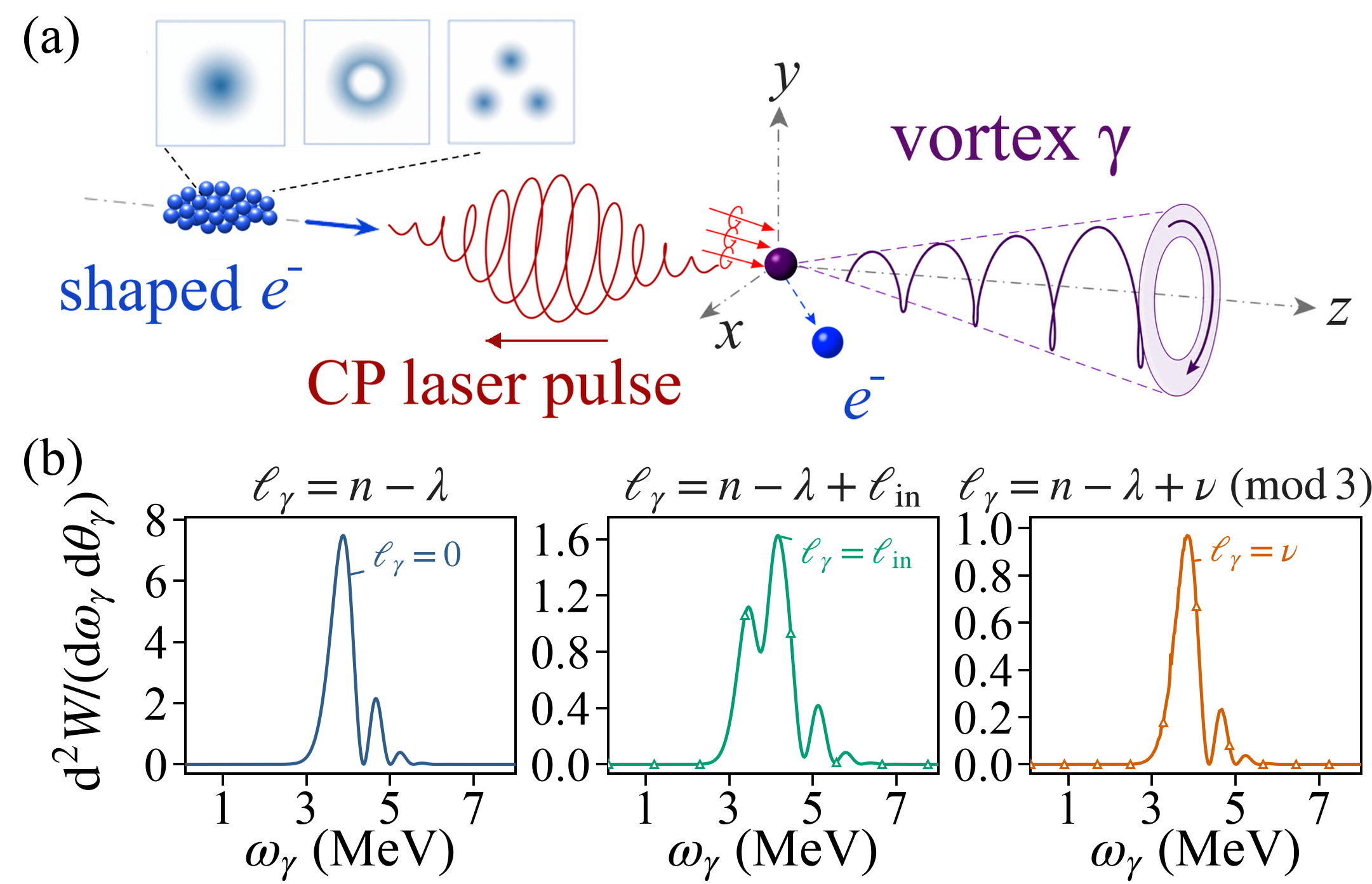}
    \caption{
Schematic illustration of electron-wavepacket control of photon orbital angular momentum in nonlinear Compton scattering.
(a) A relativistic electron with a structured transverse wave packet collides head-on with a circularly polarized laser pulse and emits vortex $\gamma$ photons, whose OAM distribution is determined by the transverse structure of the electron wave packet.
(b) Photon energy spectra generated from Gaussian (blue), vortex (green, carrying intrinsic OAM $\ell_{\rm in}$), and three-component momentum-superposition (orange, $\nu=1+3k$, $k=0,\pm1,\pm2,\dots$) electron wave packets. The corresponding photon OAM channels at the leading harmonic peak ($n=1$) for positive photon helicity ($\lambda=+1$) are indicated.
}
    \label{fig1}
\end{figure}

In this letter, we develop a general analytical framework for NCS driven by an arbitrarily shaped electron wave packet. As illustrated schematically in Fig.~\ref{fig1}(a), a structured relativistic electron collides with a circularly polarized laser pulse and emits vortex $\gamma$ photons. We derive a generalized angular-momentum selection rule demonstrating that the transverse Fourier spectrum of the electron wave packet is directly mapped onto the OAM spectrum of the emitted photons. Within this framework, Gaussian wave packets and vortex electrons emerge naturally as limiting cases, while arbitrary transverse wave-packet structures introduce additional degrees of freedom into the radiation process. 
Figure~\ref{fig1}(b) illustrates the central consequence of this principle that although the leading harmonic channel ($n=1$) dominates the photon yield in all cases, the associated photon OAM state is determined by the transverse structure of the incident electron wave packet. Our work reveals electron-wavepacket structure as a fundamental degree of freedom governing photon orbital angular momentum and establishes a general framework for structured radiation in nonlinear Compton scattering.

To establish the connection between the electron wave-packet structure and photon OAM, we consider NCS between a relativistic electron described by an arbitrary momentum-space wave packet $\rho(\mathbf p)$ and a circularly polarized laser pulse (heliciy $\lambda_0=+1$). The laser field is represented by the plane-wave four-potential
\begin{equation}
a^\mu(\tau)=\frac{m\xi}{\sqrt{2}}h(\tau)(0,\mathbf a_\perp(\tau),0),
\end{equation}
where $\xi$ is the normalized laser amplitude, $\tau=k\cdot x$ is the laser phase and $h(\tau)$ denotes the pulse envelope. 
The final-state photon and electron are projected onto Bessel states to resolve their orbital angular momenta, with the transverse kernel
\begin{equation}
a_{\kappa,\ell}(\mathbf p_\perp)
=
(-i)^\ell
\sqrt{\frac{2\pi}{\kappa}}
\delta(|\mathbf p_\perp|-\kappa)
e^{i\ell\phi_p},
\end{equation}
where $\ell$ denotes the longitudinal OAM quantum number.

After performing the transverse momentum integrations in the scattering matrix, the OAM-resolved amplitude can be expressed as
\begin{equation}
\begin{aligned}
\mathcal M_{\ell_\gamma,\ell'}
=&
\int d\chi d\phi\,
\rho(\mathbf p)
e^{-i(\ell_\gamma+\ell')\phi}
\widetilde{\mathcal M}_{\rm PW},
\end{aligned}
\end{equation}
where $\phi$ is the azimuthal angle of the incident electron transverse momentum, and $\chi$ describes the relative geometry of the final-state transverse momenta. Introducing the total OAM variable $L=\ell_\gamma+\ell'$ and the relative OAM $\ell'=r$, the amplitude can be rewritten as
\begin{equation}
\mathcal M_{L,r}
=
(-1)^r
\int d\chi e^{-ir\chi}
\int d\phi e^{-iL\phi}
\widetilde{\mathcal M}_{\rm PW}.
\end{equation}
This expression reveals that the OAM-resolved radiation amplitude is governed by the azimuthal Fourier components of the incident electron wave packet. Expanding the electron wave packet in the Bessel basis allows us to write its transverse structure as a superposition of Bessel states
\begin{equation}
	\rho_\perp(\mathbf{p}_\perp)\ket{p,\sigma} = \sum_{\nu=-\infty}^{\infty}\int d\kappa\, i^\nu \sqrt{\frac{\kappa}{2\pi}}C_\nu\ket{p_+,\kappa,\nu,\sigma}
\end{equation}
where $\nu$ denotes the OAM carried by the respective component. In the long-pulse regime, the azimuthal dependence of the plane-wave amplitude ensures that every component obeys the selection rule $\ell_\gamma = n - \lambda + \nu$ ($\lambda$ being the helicity of the emitted photon). Consequently, the scattering amplitude of a general wave packet is given by the linear superposition of the amplitudes associated with each incident electron OAM component.
\begin{equation}\label{M_sh_1}
\mathcal M^{\rm sh}_{L,r}
=
-ie
\sum_n
\int d\tilde\chi\,
\widetilde{\mathcal M}_{\rm PW}
\sum_\nu
C_\nu(p_\perp)
\delta_{\nu,L+\lambda-n}.
\end{equation}
Therefore, the conventional angular-momentum selection rule is generalized to
\begin{equation}
\delta_{L+\lambda-n,0}
\rightarrow
\sum_\nu C_\nu(p_\perp)
\delta_{\nu,L+\lambda-n},
\end{equation}
or equivalently,
\begin{equation}
\ell_\gamma=n-\lambda+\nu .
\end{equation}
This relation constitutes the central result of this work. The additional index $\nu$ is determined by the transverse Fourier structure of the electron wave packet rather than by an intrinsic electron OAM. Thus, Gaussian wave packets and vortex electrons correspond to the special cases with a single Fourier component, while arbitrary wave-packet structures introduce additional OAM channels weighted by their Fourier amplitudes $C_\nu$. The transverse electron wave packet therefore acts as a controllable degree of freedom that is directly imprinted onto the OAM spectrum of emitted photons.

To demonstrate the physical consequences of the generalized selection rule, we first consider a finite momentum-superposition state as the simplest realization of a shaped electron wave packet. This example allows the role of the wave-packet Fourier spectrum to be identified analytically and illustrates how individual photon OAM channels can be engineered.

The incident electron is prepared as a superposition of $M$ Gaussian momentum components,
\begin{equation}
\rho_M=\frac{1}{\sqrt{N_M}}\sum_{j=0}^{M-1}c_j\rho_j,
\end{equation}
where $c_j=\frac{e^{i\mu_0\phi_j}}{\sqrt{M}}, 
\rho_j=\rho_z
\exp\!\left[-\frac{\sigma_\perp^2}{2}
|\mathbf p_\perp-\mathbf K_j|^2\right]$.
The momentum components are uniformly distributed in azimuth, $
\phi_j=\phi_0+\frac{2\pi j}{M}$,
forming a regular polygon in transverse momentum space.

Substituting this wave packet into Eq.~(\ref{M_sh_1}) yields
\begin{equation}
C_\nu
=
c_K
I_\nu(z)
\sum_{j=0}^{M-1}
c_j
e^{-i\nu\phi_j},
\label{CM}
\end{equation}
with $
c_K=\frac{
e^{-\sigma_\perp^2(p_\perp^2+K^2)/2}}
{\sqrt {N_M}},
\qquad
z=\sigma_\perp^2Kp_\perp$.
Equation~(\ref{CM}) immediately reveals two independent control mechanisms. The discrete Fourier summation acts as an OAM filter, allowing only channels satisfying
\[
L+\lambda-n
=
\mu_0 + Mk \quad(k=0,\pm1,\pm2 ...),
\]
whereas the modified Bessel function determines the relative weight of each surviving channel. Consequently, the dominant emitted OAM channel is selected by the phase parameter $\nu_0$, while the population of the remaining channels can be continuously tuned through the transverse momentum scale $K$.
\begin{figure}[tbp] 
    \centering
    \includegraphics[width=0.48\textwidth]{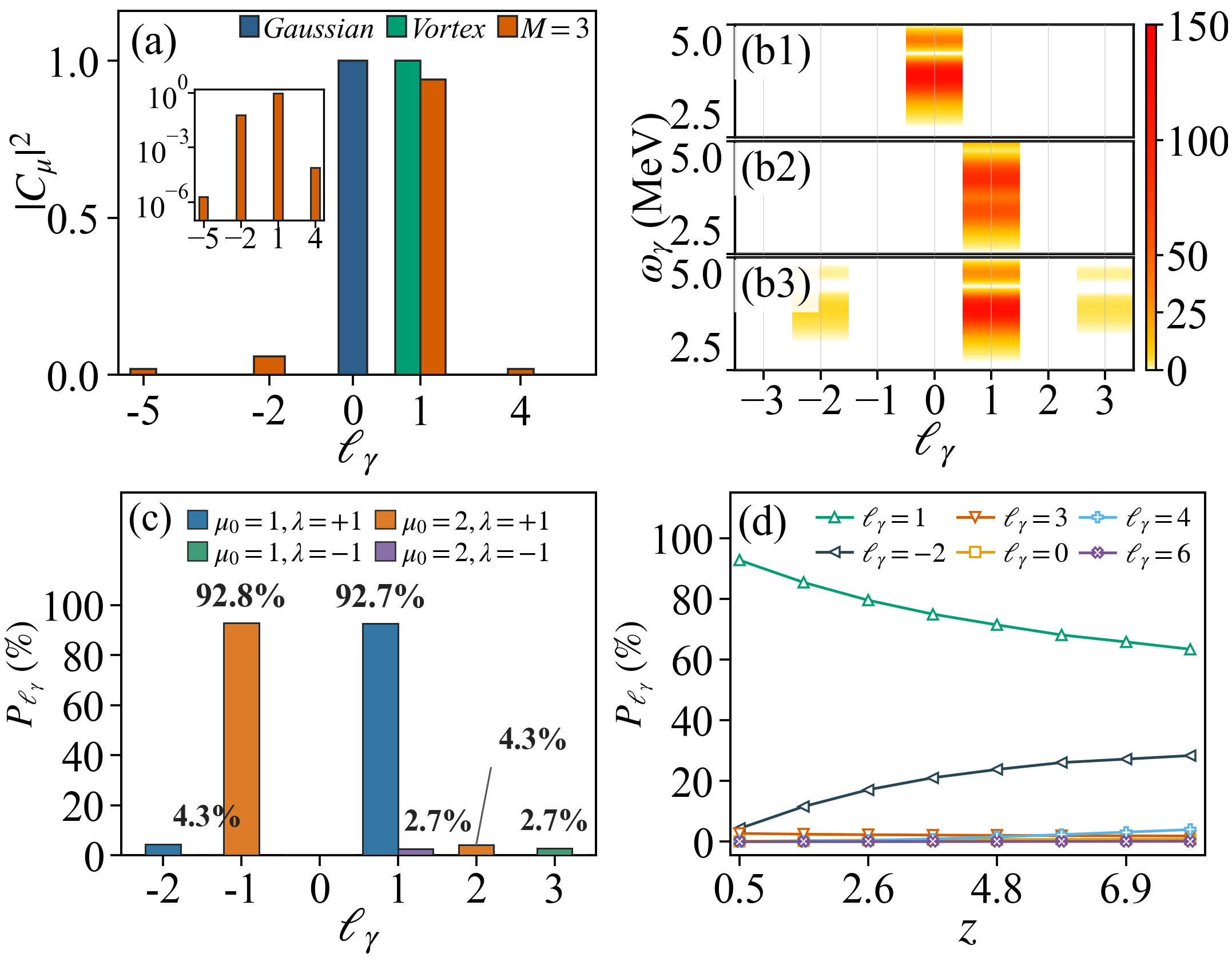}
    \caption{ (a) Fourier spectra of the transverse electron wave packets used in Fig.~1(b), including a Gaussian wave packet, a vortex electron with intrinsic OAM $\ell_{\rm in}=1$, and a three-component momentum-superposition state with $\mu_0=1$.
(b) Energy-resolved photon OAM distributions for the three electron states shown in (a). 
(c) OAM channel decomposition of photons within the energy range of $2.5-5$ MeV for momentum-superposition electron wave packets with different angular Fourier components $\mu_0$.
(d) Relative contributions of different OAM channels as a function of the transverse momentum parameter $z$ for photons with $\mu_0=1$ in (c). Laser intensity $\xi=1$, electron energy $E=511$ MeV.}
    \label{fig2}
\end{figure}

The generalized selection rule derived above establishes a direct mapping
between the transverse Fourier spectrum of an electron wave packet and
the OAM spectrum of emitted photons. For the momentum-superposition
state considered here, the modulation factor $C_\nu$ contains three
independent control parameters. The number of components $M$ determines
the spacing between accessible Fourier components, the phase index
$\mu_0$ determines their offset and therefore the dominant OAM channel,
while the parameter $z=\sigma_\perp^2 Kp_\perp$ regulates their relative
populations through the modified Bessel function $I_\nu(z)$. Specifically,
the azimuthal Fourier projection term
$\sum_{j=0}^{M-1}c_j e^{-i\mu\phi_j}$ acts as a selection filter,
allowing only the components satisfying
$\nu=\mu_0+kM$ to contribute. Therefore, the transverse structure of the
electron wave packet provides an additional degree of freedom for
controlling the OAM distribution of the emitted photons.

To demonstrate this correspondence, we compare three representative
electron states, a Gaussian wave packet, a vortex electron, and a
three-component momentum-superposition state. Their angular Fourier
spectra $|C_\nu|^2$ are shown in Fig.~\ref{fig2}(a). For a Gaussian wave
packet, only the zeroth-order Fourier component exists, $C_0\neq0$,
corresponding to the conventional unstructured electron state. A vortex
electron is characterized by a single Fourier component determined by its
intrinsic OAM, $C_{\ell_{\rm in}}\neq0$. In contrast, the
momentum-superposition state generates a series of discrete Fourier
components satisfying $\nu=\mu_0+kM$, providing multiple controllable
channels for photon OAM generation.

The corresponding energy-resolved photon OAM distributions are presented
in Fig.~\ref{fig2}(b). For all three electron states, the dominant
radiation contribution originates from the first harmonic ($n=1$), which
corresponds to the highest photon yield in nonlinear Compton scattering.
For the Gaussian wave packet [Fig.~\ref{fig2}(b1)], the conventional
angular-momentum selection rule is recovered, and the dominant photons
carry zero OAM. For the vortex electron [Fig.~\ref{fig2}(b2)], the
intrinsic electron OAM shifts the angular-momentum balance, resulting in
a dominant photon OAM channel of $\ell_\gamma=\ell_{\rm in}=1$. In
contrast, the momentum-superposition wave packet [Fig.~\ref{fig2}(b3)]
produces a nonzero photon OAM despite the absence of a definite intrinsic
electron OAM. The dominant OAM channel, $\ell_\gamma=n-\lambda-\nu=1$, is determined by the Fourier
component $\mu_0=1$ of the electron wave packet rather than by an initial
angular-momentum quantum number. This demonstrates that the OAM state of
the high-yield radiation channel is not fundamentally fixed by the
multiphoton absorption process, but can be governed by the transverse
structure of the incident electron wave packet itself.

The detailed OAM composition of the high-yield photon emission channel is further analyzed in Fig.~\ref{fig2}(c). Here, photons in the energy range
of $2.5-5$ MeV around the first harmonic peak are selected, corresponding
to the dominant contribution to the emitted photon yield. For $M=3$ and
$z=0.5$, the allowed Fourier components satisfy
$\nu=\mu_0+3k$, leading to distinct OAM distributions for different
values of $\mu_0$. For $\mu_0=1$, the dominant channel corresponds to
$\lambda=+1$ and $\ell_\gamma=1$, accounting for $92.7\%$ of the photons
within this energy range. The secondary channels,
$\lambda=+1$, $\ell_\gamma=-2$ and $\lambda=-1$, $\ell_\gamma=3$,
contribute only $4.3\%$ and $2.7\%$, respectively. These additional
channels originate from higher-order allowed Fourier components
$\nu=\mu_0+3k$ and the two possible helicity states of the emitted photon
in the angular-momentum balance.

When the phase index is changed to $\mu_0=2$, the Fourier spectrum of the
electron wave packet is shifted correspondingly, resulting in a complete
redistribution of the photon OAM spectrum. The dominant channel becomes
$\lambda=+1$, $\ell_\gamma=-1$, contributing $92.8\%$, while the secondary
channels $\lambda=+1$, $\ell_\gamma=2$ and
$\lambda=-1$, $\ell_\gamma=1$ contribute $4.3\%$ and $2.7\%$,
respectively. Importantly, this OAM switching occurs without modifying
the photon-energy spectrum or the driving laser field. Therefore, the
OAM purity of the dominant high-yield $\gamma$-ray emission channel can
be directly engineered through the transverse Fourier structure of the
electron wave packet.

While $\mu_0$ determines the position of the dominant OAM channel, the
parameter $z$ controls the relative weights of the allowed channels.
Figure~\ref{fig2}(d) shows the evolution of the OAM composition. As $z$ increases from 0.5 to 0.8,
the contribution of the dominant channel
$\lambda=+1$, $\ell_\gamma=1$ decreases from approximately $92.8\%$ to
$63.4\%$, while the secondary channel
$\lambda=+1$, $\ell_\gamma=-2$ increases from $4.3\%$ to about $28.4\%$.
Other accessible OAM channels remain below $4\%$. This continuous
redistribution originates from the $z$ dependence of the Fourier
coefficients through $I_\mu(z)$ and demonstrates that electron
wave-packet engineering provides simultaneous control over both the
selected OAM channel and its modal purity.

The momentum-superposition state discussed above provides a simple example for revealing the mapping between the electron-wavepacket Fourier spectrum and the photon OAM distribution. We next extend the analysis to another representative class of electron states generated through transverse phase shaping, which is closely related to recent developments in free-electron quantum optics\cite{garcia2021optical,chirita2022transverse}. The outgoing electron wave packet is written as
\[
\Psi_{\rm out}(r,\theta)
=
\Psi_{\rm in}(r,\theta)e^{i\varphi(r,\theta)},
\]
where the transverse phase profile is chosen as
\[
\varphi(r,\theta)
=
a\sin(M\theta)+b\sin(2M\theta).
\]
Following the generalized selection rule, the angular Fourier components generated by this phase modulation are given by
\begin{equation}\label{dsin}
C_{sM}
=
\mathcal{N}d_s(a,b)H_{sM}(p_\perp\sigma_\perp),
\end{equation}
where $d_s(a,b)=\sum_m J_{s-2m}(a)J_m(b)$ determines the angular Fourier weights induced by the phase modulation, while $H_{sM}=\int_0^\infty dt\, e^{-t}J_{sM}(p_\perp\sigma_\perp \sqrt{2t})$ accounts for the transverse momentum dependence. Therefore, this example provides a more general demonstration that arbitrary transverse electron structures can be directly encoded into the photon OAM spectrum. Consequently, the generalized selection rule is modified as
\[
\ell_\gamma=n-\lambda-sM .
\]

\begin{figure}[tbp] 
    \centering
 \centering
    \includegraphics[width=0.48\textwidth]{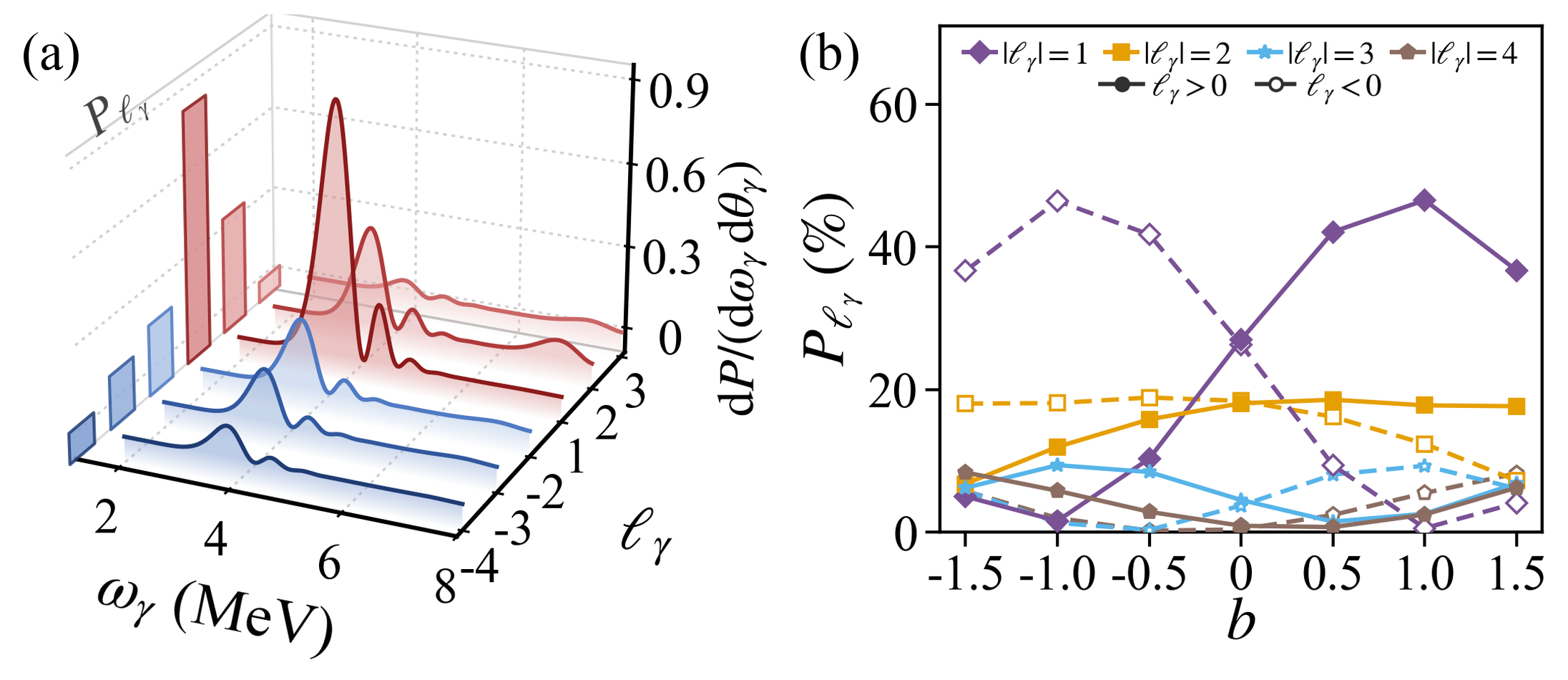}    
    \caption{(a) Energy-resolved photon OAM spectrum for the phase-shaped electron wave packet with shaping parameters $a=2.4$ and $b=1.0$. The horizontal axes represent the photon energy and OAM quantum number $\ell_\gamma$, while the vertical axis denotes the emission probability. (b) Relative contributions of different photon OAM channels within the first-harmonic energy range ($2.5-5$ MeV) as a function of the phase-shaping parameter $b$. Laser intensity $\xi=1$, electron energy $E=511$ MeV.}
    \label{fig3}
\end{figure}

Figure~\ref{fig3}(a) shows the energy-resolved photon OAM distribution generated by a phase-shaped electron wave packet with $a=2.4$ and $b=1.0$. Different from the vortex-electron case, this electron state does not possess a definite intrinsic OAM. Nevertheless, multiple photon OAM channels are generated simultaneously, and all of them exhibit their maximum emission probability around the first harmonic peak ($n=1$). Selecting photons in the high-yield energy range of $2.5-5$ MeV, the dominant contribution comes from the $\ell_\gamma=1$ channel, accounting for about $46.5\%$ of the total emission, followed by the $\ell_\gamma=2$ channel with approximately $17.8\%$. The remaining channels, including $\ell_\gamma=-2$, $\ell_\gamma=\pm3$, and $\ell_\gamma=\pm4$, contribute smaller fractions of the radiation. This result demonstrates that a continuous phase-shaped electron wave packet can generate structured $\gamma$ radiation without requiring a predefined electron angular momentum state.

The tunability of the photon OAM distribution is further demonstrated in Fig.~\ref{fig3}(b), where the relative contributions of different OAM channels within the first-harmonic energy range ($2.5-5$ MeV) are shown as a function of the phase parameter $b$ with fixed $a=2.4$. According to Eq.~(\ref{dsin}), the relative Fourier weights are determined by the ratio between the two phase-modulation components, and the symmetry of the phase profile leads to a symmetric redistribution of the OAM channels with respect to $b=0$. As $b$ increases from $0$ to $1.5$, the contribution of the $\ell_\gamma=1$ channel first increases from about $26.9\%$ to nearly $46.5\%$ and then decreases, while the $\ell_\gamma=-1$ channel is initially suppressed and subsequently increases to about $4.1\%$. Meanwhile, the $\ell_\gamma=2$ channel remains nearly unchanged, whereas the contribution of the $\ell_\gamma=-2$ channel decreases from approximately $18.4\%$ to $7.2\%$. Other accessible channels exhibit only moderate variations and remain below $9.5\%$. These results confirm that the OAM composition of high-yield vortex $\gamma$ photons can be continuously controlled by tailoring the transverse phase structure of the electron wave packet.

In summary, we have established a general analytical framework for nonlinear Compton scattering driven by arbitrarily shaped electron wave packets and uncovered a direct connection between the quantum structure of an electron and the orbital angular momentum of emitted photons. The derived generalized angular-momentum selection rule, $
\ell_\gamma=n-\lambda+\nu$, demonstrates that the transverse Fourier spectrum of an electron wave packet is directly encoded into the photon OAM distribution, revealing electron-wavepacket structure as an intrinsic degree of freedom in structured radiation generation. Beyond the conventional picture based on angular-momentum transfer from initial particles and electromagnetic fields, our results show that the internal structure of a quantum emitter itself can determine the angular momentum distribution of its radiation. This finding provides a general principle for understanding and manipulating structured photons in high-energy quantum processes.

\textit{Acknowledgements-}The authors thank K. Z. Hatsagortsyan for helpful discussions. This work is supported by the Scientific Research Innovation Capability Support Project for Young Faculty (Grant No. ZYGXQNJSKYCXNLZCXM-E12) and the National Natural Science Foundation of China (Grants Nos. 12574346 and 12222507).

\bibliography{reference}

\end{document}